\documentclass[12pt]{article}

\usepackage{amssymb}
\usepackage{amsmath}
\usepackage{amsfonts}

\newcommand{\R}{\mathbb{R}}
\newcommand{\C}{\mathbb{C}}

\newcommand{\be}{\begin{equation}}
\newcommand{\bea}{\begin{eqnarray}}
\newcommand{\eea}{\end{eqnarray}}

\newcommand{\kt}{\rangle}
\newcommand{\br}{\langle}

\newcommand{\ed}{\end{document}}

\begin{document}

\title{A Critique of ${\cal PT}$-Symmetric Quantum Mechanics}
\author{\\
Ali Mostafazadeh\thanks{E-mail address: amostafazadeh@ku.edu.tr}\\
\\ Department of Mathematics, Ko\c{c} University,\\
Rumelifeneri Yolu, 34450 Sariyer,\\
Istanbul, Turkey}
\date{ }
\maketitle
\begin{abstract} We study the physical content of the
${\cal PT}$-symmetric complex extension of quantum mechanics as
proposed in Bender {\em et al}, Phys.\ Rev.\ Lett.\ {\bf 80}, 5243
(1998) and {\bf 89}, 270401 (2002), and show that as a fundamental
probabilistic physical theory it is neither an alternative to nor
an extension of ordinary quantum mechanics. We demonstrate that
the definition of a physical observable given in the above papers
is inconsistent with the dynamical aspect of the theory and offer
a consistent notion of an observable.
\end{abstract}
%\vspace{.3cm} PACS numbers: 11.30.Er, 03.65-w, \vspace{5mm}

\baselineskip=24pt

The past five years have witnessed a great deal of research
activity on the subject of ${\cal PT}$-symmetric quantum
mechanics. This was mainly initiated by Bender and Boettcher's
demonstration \cite{bender1} that the spectrum of the
Hamiltonians:
    \be
    \hat H=\hat p^2+\hat x^2(i\hat x)^\nu~~~~~{\rm with}~~~~~\nu\in\R^+,
    \label{H}
    \end{equation}
was actually real, positive, and discrete. Here the operators
$\hat p$ and $\hat x$ are defined according to $(\hat p\psi)(x)=
-i\partial_x\psi(x)$, $(\hat x\psi)(x)=x\psi(x)$, and for $\nu\geq
2$ the eigenvalue problem for $\hat H$ is defined by imposing
vanishing boundary conditions on an appropriate contour $C$ in the
complex $x$-plane \cite{bender1}. The interest in the properties
of the Hamiltonians~(\ref{H}) was boosted by the more recent
findings of Bender, Brody, and Jones \cite{bender3} who showed
that a (unitary) probabilistic formulation of ${\cal
PT}$-symmetric quantum mechanics based on the
Hamiltonians~(\ref{H}) was possible.

The initial announcement of the results regarding the reality of
the spectrum of the Hamiltonians (\ref{H}) led to a great deal of
surprise and controversy, for such a Hamiltonian was apparently
``non-Hermitian''. Bender and his collaborators \cite{bender1}
argued that the unusual spectral properties of these Hamiltonians
was due to their ${\cal PT}$-symmetry: $[H,{\cal PT}]=0$, where
${\cal P}$ and ${\cal T}$ stand for the ``parity'' and
``time-reversal'' operators defined by $(P\psi)(x):=\psi(-x)$ and
$(T\psi)(x):=\psi(x)^*$, respectively. They also showed that the
eigenfunctions $\phi_n$ of $H$, i.e., the solutions of the
Schr\"odinger equation
    \be
    \left[-\frac{d^2}{d^2x}+x^2(ix)^\nu\right]\phi_n(x)=E_n\phi_n(x)
    \label{sch-eq}
    \end{equation}
fulfilling vanishing boundary conditions on the contour $C$ were
orthogonal with respect to both the indefinite ${\cal PT}$-inner
product: $(\psi,\phi):=\int_C dx~[{\cal PT}\psi(x)]\phi(x)$, and
the positive-definite ${\cal CPT}$-inner product \cite{bender3}:
    \be
    (\psi,\phi)_+:=\int_C dx~[{\cal CPT}\psi(x)]\phi(x),
    \label{CPT-inn}
    \end{equation}
where ${\cal C}$ is the so-called charge-conjugation operator
defined through its ``position-representation'' according to
${\cal C}(x,y)=\sum_n\phi_n(x)\phi_n(y)$.

The main assertion of \cite{bender3} is that not only the ${\cal
PT}$-symmetric Hamiltonians~(\ref{H}) can be used to define the
real energy levels of a quantum system, but that they are also
capable of defining a unitary time-evolution provided that one
adopts the ${\cal CPT}$-inner product~(\ref{CPT-inn}) to define a
Hilbert space structure on the space of state vectors. It is this
latter observation that has raised the expectations of a number of
theoretical physicists to consider the ${\cal PT}$-symmetric
quantum mechanics as a possible alternative to or an extension of
the ordinary quantum mechanics (QM). The aim of this letter is to
show that these expectations do not have a valid ground and that
indeed a consistent probabilistic ${\cal PT}$-symmetric quantum
theory is doomed to reduce to ordinary QM.

We begin our analysis by elaborating on a number of ambiguities
that have so far obscured the physical content of the ${\cal
PT}$-symmetric quantum mechanics.

It is often claimed that unlike Hermiticity the ${\cal
PT}$-symmetry is a physical requirement, for it means symmetry
under space-time reflections \cite{bender3}. This argument rests
on the assumption (illegally imported from the ordinary QM) that
the variable $x$ appearing in Eq.~(\ref{sch-eq}) is to be
associated with the position of a particle, i.e., a point in the
physical configuration space. In QM the argument $x$ of a wave
function does not have an a priori physical meaning. It is merely
a (generalized) eigenvalue of a linear operator $\hat x$. $x$
inherits a physical meaning from $\hat x$ which represents a
physical observable called the position. This in turn relies on
the postulates that in QM physical observables are self-adjoint
(Hermitian) operators acting in the space of state vectors of the
system, that the latter is a separable Hilbert space (or an
appropriately extended Dirac space), and that there is a
well-defined measurement theory that describes how one should
associate the theoretical predictions with the experimental data.
Even these alone are not sufficient to relate the variable $x$ to
a point in the physical space unless one enforces Dirac's
canonical quantization program and makes a connection between the
observable $\hat x$ and the position of the corresponding
classical particle.

This argument shows that before being able to identify the values
of the variable $x$ of Eq.~(\ref{sch-eq}) with points of the
physical space or viewing ${\cal P}$ as the usual parity
(space-reflection) operator, one must first formulate the ${\cal
PT}$-symmetric quantum mechanics as a physical theory. This
entails addressing the following questions: 1.~What is the
mathematical nature of the space of state vectors? 2.~What are the
observables? 4.~How are the observables measured? 4.~How does the
theory relate to known theories? Alternatively, is there a
correspondence principle that would, for example, justify calling
the operator $\hat x$ of Eq.~(\ref{H}) the ``position operator''.

As far as the results of Refs.~\cite{bender1,bender3} are
concerned, $x$ is just the independent variable of the solutions
of a certain eigenvalue problem for the differential operator
$\hat H$. Nevertheless, it is a fact that this eigenvalue problem
defines an associated complex vector space ${\cal V}$, namely the
span of the eigenfunctions of $\hat H$, and that this vector space
can be endowed with an appropriate (positive-definite) inner
product and made into a separable Hilbert space (through Cauchy
completion \cite{reed-simon}). It is important to note that before
constructing this Hilbert space one cannot decide whether the
operator $\hat H$ is Hermitian or not. This also raises the issue
of certain ambiguity in the terminology used in
\cite{bender1,bender3}.

Often, one uses the terms ``Hermitian'' and ``self-adjoint''
synonymously for operators $A$ that act in a Hilbert space ${\cal
K}$ and satisfy the defining relation:
    \be
    \br \psi,A\phi\kt=\br A\psi,\phi\kt,
    \label{self-adj}
    \end{equation}
where $\psi,\phi\in{\cal K}$ are arbitrary and $\br\cdot,\cdot\kt$
denotes the inner product of ${\cal K}$. It is a simple result of
linear algebra that the matrix representation of a self-adjoint
operator $A$ in an orthonormal basis is a Hermitian matrix, i.e.,
if $A$ satisfies (\ref{self-adj}), then the matrix elements
$A_{ij}$ of $A$ in any orthonormal basis satisfy
$A_{ij}=A_{ji}^*$. This is the origin of the use of the term
``Hermitian'' for a self-adjoint operator $A$ (that by definition
satisfies (\ref{self-adj}).) Apart from the technical issues
related to the domain of $A$, there is no danger of using this
terminology in ordinary QM where the Hilbert space has a fixed
inner product and one works with orthonormal bases that are
usually formed out of the eigenvectors of the relevant (commuting)
observables.

The situation is quite different in ${\cal PT}$-symmetric QM where
a priori neither the inner product nor the observables are fixed.
As a result, one must refrain from calling the
Hamiltonians~(\ref{H}) ``non-Hermitian'' and referring to ${\cal
C}(x,y)$ as the ``position representation'' of the operator ${\cal
C}$. In fact, as the general results reported in \cite{p2} and the
particular constructions given in \cite{bender3} show, the
Hamiltonians~(\ref{H}) are ``Hermitian'' with respect to some
inner product after all.

The use of the term ``non-Hermitian'' for these and similar
Hamiltonians \cite{bender1,bender3} stem from a rather naive
definition of a Hermitian operator according to which a linear
operator $A$ is called a ``Hermitian operator'' if its matrix
representation in the $x$-representation is a Hermitian (infinite)
matrix, $A(x,y)^*=A(y,x)$. Unlike the definition of a Hermitian
operator based on the condition~(\ref{self-adj}), this definition
suffers from the fact that it is manifestly basis-dependent. In
fact, it is not difficult to see that the operator $\hat x$ is not
self-adjoint with respect to the ${\cal CPT}$-inner product.
Therefore, the $x$-basis is not an orthonormal basis and the
above-mentioned correspondence between the Hermiticity
(self-adjointness) of an operator (as defined by (\ref{self-adj}))
and the Hermiticity of its matrix representation does not hold in
the $x$-representation. This shows that taking $A(x,y)^*=A(y,x)$
as the definition of a ``Hermitian operator'' is quite misleading.

The same problem arises when one defines the notion of a
``symmetric operator'' \cite{bender3} through the requirement that
the matrix elements of an operator $A$ in the $x$-representation
form a symmetric matrix, $A(x,y)=A(y,x)$. In particular, the
proposal of identifying physical observables with ``symmetric''
${\cal CPT}$-invariant linear operators outlined in \cite{bender3}
is ill-defined unless a prescription is provided that fixes a
basis so that one can determine whether an operator admits a
symmetric matrix representation in this basis. As we argued above
the choice of the $x$-representation used in \cite{bender3} cannot
be motivated by physical considerations.

Next, we wish to return to the problem of determining the physical
observables for the ${\cal PT}$-symmetric quantum systems defined
by the Hamiltonians~(\ref{H}). Having used the inner product
(\ref{CPT-inn}) to endow the vector space ${\cal V}$ with the
structure of a Hilbert space ${\cal H}$, we ensure that the
time-evolution determined by the time-dependent Schr\"odinger
equation, $i\partial_t\psi(t)=\hat H\psi(t)$, is unitary. If we
wish to adopt the same measurement theory as in the ordinary QM,
we are forced to define the observables as Hermitian operators
acting in ${\cal H}$. This definition differs from the one given
in \cite{bender3} that identifies the observables with symmetric
${\cal CPT}$-invariant operators. We will first explore the
consequences of our definition of the observables and then show
using a finite-dimensional toy model that indeed the definition
given in \cite{bender3} is inconsistent.

The identification of the observables with Hermitian operators
acting in the Hilbert space ${\cal H}$ immediately excludes the
operators $\hat x$ and $\hat p$ as candidates for physical
observables, for unlike the Hamiltonian $\hat H$ they fail to be
Hermitian with respect to the inner product~(\ref{CPT-inn}) of
${\cal H}$. In particular, they cannot be identified as the
quantum analogs of the position and momentum of a classical
particle. This in turn leads to the natural question: What are the
position and momentum operators in the ${\cal PT}$-symmetric
quantum mechanics?

In order to respond to this question, we recall a well-known
mathematical result about Hilbert spaces, namely that {\em up to
isomorphisms there is a unique infinite-dimensional separable
Hilbert space} \cite[Theorem II.7]{reed-simon}. This implies that
${\cal H}$ may be mapped onto the Hilbert space $L^2(\R)$ by a
unitary linear transformation, i.e., there exists a linear map
${\cal U}:{\cal H}\to L^2(\R)$ satisfying
    \be
    (\psi,\phi)_+=\br{\cal U}\psi|{\cal U}\phi\kt,
    \label{unitary}
    \end{equation}
where $\psi,\phi\in{\cal H}$ are arbitrary and $\br\cdot|\cdot\kt$
is the usual $L^2$-inner product.

Equation~(\ref{unitary}) suggests a direct method of constructing
physical observables for the ${\cal PT}$-symmetric quantum systems
associated with the Hamiltonians~(\ref{H}). These have the general
form $\hat O={\cal U}^{-1}\hat o \;{\cal U}$ where $\hat o$ is a
Hermitian operator acting in $L^2(\R)$. For example the position
and momentum observables are respectively described by $\hat
X={\cal U}^{-1}\hat x\;{\cal U}$ and $\hat P={\cal U}^{-1}\hat
p\;{\cal U}$ where $\hat x$ and $\hat p$ are the position and
momentum operators of ordinary QM.

The uniqueness of the Hilbert space structure has another
important consequence: One can describe the ${\cal PT}$-symmetric
systems defined by the Hilbert space ${\cal H}$, the Hamiltonian
$H$, and the observables $\hat O$ in terms of an ordinary quantum
system having $L^2(\R)$ as the Hilbert space, $\hat h:={\cal
U}\hat H {\cal U}^{-1}$ as the Hamiltonian, and $\hat o={\cal
U}\hat O\,{\cal U}^{-1}$ as the observables. In this sense the
${\cal PT}$-symmetric QM actually reduces to the ordinary QM.
Therefore the claim that it is a fundamental physical theory
extending QM is not valid.

The operator ${\cal U}$ used in the above discussion is in general
nonlocal: The corresponding similarity transformation maps
differential operators to nonlocal (non-differential) operators.
In particular, the Hamiltonian $\hat h$ will most probably not
have the standard (kinetic $+$ potential) form. The same nonlocal
behavior is also shared by the observables $\hat O$, in general,
and by the position and momentum operators $\hat X$ and $\hat P$,
in particular.

Because a closed-form expression for the operator ${\cal C}$ that
enters in the expression for the inner product (\ref{CPT-inn}) is
not available, one cannot obtain the explicit form of the operator
${\cal U}$ and consequently the Hamiltonian $\hat h$ and the
observables $\hat X$ and $\hat P$. One may however attempt to
obtain approximate expressions for the latter using the series
expansion method outlined in \cite{bender2}. As pointed out in
\cite{bender4} there are finite-dimensional analogs of the ${\cal
PT}$-symmetric Hamiltonians~(\ref{H}) where the operator ${\cal
C}$ is represented by a simple matrix. For these systems one may
use the machinery of the theory of pseudo-Hermitian operators
\cite{p1,p2,p3,p7} to construct the operators ${\cal U}$ and $\hat
h$ explicitly \cite{jpa-03}. In the following, we shall offer an
independent re-examination of a two-dimensional model introduced
in \cite{bender3} to provide a concrete realization of our general
results and to demonstrate the shortcomings of the definition of
the observables given in \cite{bender3}.

Consider the matrix Hamiltonian \cite{bender3}
    \be
    H:=\left(\begin{array}{cc}
    r\,e^{i\theta}&s\\
    s& r\,e^{-i\theta}\end{array}\right),
    \label{2-level}
    \end{equation}
where $r,s,\theta\in\R$, $s\neq 0$, and $|r\sin(\theta)/s|<1$. Let
${\cal T}$ be the operation of complex-conjugation of vectors
$\psi\in\C^2$, ${\cal P}:=\sigma_1$, and ${\cal
C}:=\sec\alpha~\sigma_1+i\tan\alpha~\sigma_3$, where $\sigma_k$
are the Pauli matrices:
    \be
    \sigma_1:=\left(\begin{array}{cc}
    0 & 1\\
    1 & 0\end{array}\right),~~~~~~
    \sigma_2:=\left(\begin{array}{cc}
    0 & -i\\
    i & 0\end{array}\right),~~~~~~
     \sigma_3:=\left(\begin{array}{cc}
    1 & 0\\
    0 & -1\end{array}\right),
    \label{pauli}
    \end{equation}
and $\alpha\in(-\pi/2,\pi/2)$ is defined by
$\sin\alpha=r\sin\theta/s$. Then as shown in \cite{bender3}, $H$
is ${\cal PT}$- and ${\cal C}$-symmetric and has real eigenvalues:
$\epsilon_\pm=r\cos\theta\pm s\cos\alpha$. Furthermore, $H$ is
Hermitian with respect to the ${\cal CPT}$-inner product:
    \be
    (\psi,\phi)_+:=({\cal CPT}\psi)\cdot\phi,
    \label{cpt-2}
    \end{equation}
where $\psi,\phi\in\C^2$, a dot means the ordinary dot product:
$\psi\cdot\phi:=\sum_{i=1}^2\psi_i\phi_i$, and the subscript $i$
labels the components of the corresponding two-dimensional complex
vector. Clearly, $H$ is a two-dimensional analog of the ${\cal
PT}$-symmetric Hamiltonians (\ref{H}).

According to \cite{bender3}, the Hilbert space ${\cal H}$ of the
${\cal PT}$-symmetric Hamiltonian~(\ref{2-level}) is obtained by
endowing $\C^2$ with the ${\cal CPT}$-inner product, and the
physical observables are symmetric matrices commuting with ${\cal
CPT}$. The latter have the general form
    \be
    {\cal O}=a\sigma_0+(b+c\sin\alpha)\sigma_1+(c-b\sin\alpha)
    \sigma_3.
    \label{obs}
    \end{equation}
where $\sigma_0$ stands for the $2\times 2$ identity matrix and
$a,b,c$ are arbitrary real parameters. Note that here the problem
of the basis-dependence of the notion of a ``symmetric operator''
has been avoided, because a choice for a basis of $\C^2$, namely
the standard basis $\{(\mbox{\tiny$\begin{array}{c}1\\0
\end{array}$}),(\mbox{\tiny$\begin{array}{c}0\\1\end{array}$})\}$,
is made and all the relevant operators are described by their
matrix representations in this basis. The latter is however not an
orthonormal basis with respect to the ${\cal CPT}$-inner
product~(\ref{cpt-2}), and there is no physical reason for
choosing it over the other bases of $\C^2$.

Next, we give an equivalent quantum description of the above
system using a self-adjoint Hamiltonian $h$ acting in the ordinary
two-dimensional Hilbert space $\C^2$ endowed with the Euclidean
inner product: $\br\psi|\phi\kt:=\sum_{i=1}^2\psi_i^*\phi_i$. We
shall denote this Hilbert space by ${\cal E}$.

Introducing
    \be
    {\cal U}:=\frac{1}{\sqrt{2\cos\alpha}}
    \left(\begin{array}{cc}
    e^{i\alpha/2}&e^{-i\alpha/2}\\
    -i e^{i\alpha/2}& ie^{i\alpha/2}\end{array}\right),
    \label{U=}
    \end{equation}
we can easily check that indeed for all $\psi,\phi\in\C^2$,
$(\psi,\phi)_+=\br{\cal U}\psi|{\cal U}\phi\kt$, i.e., ${\cal U}$
is a unitary operator mapping ${\cal H}$ onto ${\cal E}$.
Furthermore, we have
    \be
    h:={\cal U}H{\cal U}^{-1}=\left(\begin{array}{cc}
    \epsilon_+&0\\
    0& \epsilon_-\end{array}\right)=
    r\cos\theta~\sigma_0+s\cos\alpha~\sigma_3,
    \label{h=}
    \end{equation}
which is clearly a self-adjoint operator acting in ${\cal E}$. It
describes the dipole interaction of a nonrelativistic spin-half
particle with a magnetic field aligned along the $z$-axis. The
observables of this system are simply the spin operators
$s_\mu=\sigma_\mu/2$ and their linear combinations with real
coefficients.

In our formulation, the observables $O$ associated with the ${\cal
PT}$-symmetric description $({\cal H},H)$ of the above system are
obtained from the observables $o$ of the ordinary quantum
description $({\cal E},h)$ via the similarity transformation
defined by ${\cal U}$. They have the general form:
    \be
    O=\sum_{\mu=0}^3 c_\mu~S_\mu,
    \label{obs-3}
    \end{equation}
where $c_\mu$ are real constants and $S_\mu:={\cal
U}^{-1}s_\mu\,{\cal U}$. More specifically
    \be
    S_0=\frac{1}{2}\,\sigma_0,~~~~S_1=-\frac{1}{2}\,\sigma_2,~~~~
    S_2=\frac{1}{2}(\tan\alpha~\sigma_1-\sec\alpha~\sigma_3),~~~~
    S_3=\frac{1}{2}(\sec\alpha~\sigma_1+\tan\alpha~\sigma_3).
    \label{S}
    \end{equation}

The physical system described by the ${\cal PT}$-symmetric
Hamiltonian $H$, the Hilbert space ${\cal H}$, and the observables
$O$ can be more conveniently described in terms of the Hamiltonian
$h$, the Hilbert space ${\cal E}$, and the observables
$o=\sum_{\mu=0}^3 c_\mu~s_\mu$. As we mentioned above, this system
corresponds to the dipole interaction of a nonrelativistic
spin-half particle with a magnetic field.

It is not difficult to see that indeed the observables~(\ref{obs})
constitute a three-parameter subfamily of the observables
(\ref{obs-3}); the former are linear combinations of $S_0,S_2$ and
$S_3$ with real coefficients. This shows that at least for the
model~(\ref{2-level}), our definition of an observable is more
general than the one given in \cite{bender3}. We will next show
that indeed the latter is physically unacceptable and that it
leads to an explicit inconsistency in the Heisenberg picture.

The restriction that the observables must be symmetric (or ${\cal
CPT}$-invariant) matrices excludes the possibility of measuring
$S_1$. In physical terms it means that one cannot measure the spin
of the particle along the $x$-axis. This is clearly an artificial
restriction without any physical justification. It is also in
conflict with the isotropy of the Euclidean space. Relaxing the
(basis-dependent) requirement of symmetry and defining the
observables as arbitrary ${\cal CPT}$-invariant operators is also
not viable for it violates the condition that the observables must
have a real spectrum.

For the ${\cal PT}$-symmetric system defined by the
Hamiltonian~(\ref{2-level}), the Heisenberg-picture observables
$O_H$ are related to the Schr\"odinger-picture observables $O$
according to $O_H(t)=e^{itH}Oe^{-itH}$. Now, let $O=S_2$ which
according to (\ref{S}) and (\ref{obs}) is both symmetric and
${\cal CPT}$-invariant, i.e., it is an observable in the sense of
\cite{bender3}. One can use the similarity transformations that
relate $H$ and $S_2$ to $h$ and $s_2$ and the properties of the
Pauli matrices to compute the Heisenberg-picture observable
associated with $S_2$. This yields
    \be
    O_H(t)=\sin[2s\cos(\alpha)t]~S_1+\cos[2s\cos(\alpha)t]~S_2.
    \label{S2}
    \end{equation}
Because $S_1$ is neither symmetric nor ${\cal CPT}$-invariant, so
is $O_H(t)$ (except when $t$ is an integer multiple of
$\pi/(2s\cos\alpha)$ and $O_H(t)=\pm S_2$.) This shows that
defining observables as symmetric ${\cal CPT}$-invariant operators
is inconsistent with the dynamical aspects of the theory; an
observable becomes non-observable as soon as one turns on the
dynamics! Our definition of observables as Hermitian operators
acting in ${\cal H}$ does not suffer from such an inconsistency.

In summary, if one describes a physical system by a ${\cal
PT}$-symmetric Hamiltonian that generates a unitary time-evolution
and if one enforces the standard rules of quantum measurement
theory to acquire the information about the associated physical
quantities, then one can devise an ordinary quantum mechanical
description of the same system. In this sense, there is no
physical motivation for developing ${\cal PT}$-symmetric quantum
mechanics. However, in the passage from the ${\cal PT}$-symmetric
to ordinary quantum description a standard Hamiltonian is
generally mapped to a nonlocal operator while the opposite is true
for the position and momentum operators.

In spite of the difficulties with viewing ${\cal PT}$-symmetric
quantum mechanics as a genuine extension of ordinary QM, one must
admit that its study has been quite rewarding. For instance, it
has raised some fundamental issues such as the possibility of an
essentially dynamical determination of the inner product of the
Hilbert space \cite{p1,p2,bender3} and the formulation of the
theory of pseudo-Hermitian Hamiltonians \cite{p1,p2,p3,p7} that
has applications in relativistic quantum mechanics \cite{rqm} and
quantum cosmology \cite{cqg}. A proper understanding of the role
of ${\cal PT}$-symmetric Hamiltonians in
effective/phenomenological theories \cite{hatano-nelson} and
especially the physical significance of the exceptional spectral
points \cite{except} arising due to the spontaneous break-down of
${\cal PT}$-symmetry awaits further study.
$$
***************************
$$

This work has been supported by the Turkish Academy of Sciences in
the framework of the Young Researcher Award Program
(EA-T$\ddot{\rm U}$BA-GEB$\dot{\rm I}$P/2001-1-1).

%\newpage

\ed